\documentclass[10pt, conference, letterpaper]{IEEEtran}

\usepackage{graphicx}
\usepackage{minted}
\usepackage{listings}
\usepackage{xcolor}
\usepackage{multicol}
\usepackage{array}
\usepackage[utf8]{inputenc}
\usepackage[spaces,hyphens]{url}
\usepackage{caption}
\usepackage{subcaption}

\definecolor{codegreen}{rgb}{0,0.6,0}
\definecolor{codegray}{rgb}{0.5,0.5,0.5} 
\definecolor{codepurple}{rgb}{0.58,0,0.82}
\definecolor{codeblue}{rgb}{0,0,1}
\definecolor{backcolour}{rgb}{1,1,1}

\lstdefinestyle{mystyle}{
    backgroundcolor=\color{backcolour},   
    commentstyle=\color{codeblue},
    keywordstyle=\color{codeblue},
    numberstyle=\tiny\color{codegray},
    stringstyle=\color{codepurple},
    basicstyle=\ttfamily\footnotesize,
    breakatwhitespace=false,         
    breaklines=true,                 
    captionpos=b,                    
    keepspaces=true, 
    numbers=none,                    
    numbersep=5pt,                  
    showspaces=false,                
    showstringspaces=false,
    showtabs=false,                  
    tabsize=2
}

\newcommand{\shiftleft}[2]{\makebox[0pt][r]{\makebox[#1][l]{#2}}}

\graphicspath{{./}}

\begin{document}

\title{Exploiting Page Faults for Covert Communication}

\author{Sathvik Swaminathan\\
BITS Pilani\\
sathvikswaminathan@gmail.com
}

\maketitle

\begin{abstract}
We present a novel mechanism to construct a covert channel based on page faults. A page fault is an event that occurs when a process or a thread tries to access a page of memory that is not currently mapped to its address space. The kernel typically responds to this event by performing a context switch to allow another process or thread to execute while the page is being fetched from the disk. We exploit this behavior to allow a malicious process to leak secret data to another process, bypassing the isolation mechanisms enforced by the operating system. These attacks do not leverage timers and are hardware-agnostic. Experimental results demonstrate that this attack can achieve a bit error rate of under {4\%}. 
\end{abstract}

\section{Introduction}
A covert channel is an intentional communication between two
parties, a trojan, and a spy process, through a medium that was not designed to be a communication channel. The trojan and spy coordinate to transfer information by inducing and perceiving changes in the state of shared resources. Modern processors share several resources, such as caches, memory, branch predictors, and libraries, all of which introduce a variety of covert and side-channel attacks \cite{e, f, k, l, n, o}. This paper introduces a novel mechanism for covert communication based on page faults. This attack works entirely without timers, utilizing system calls (\textit{posix\_fadvise}\cite{j}) to manipulate the state of the page cache. This attack exploits a primary performance-enhancing mechanism in the operating system known as a context switch. 

A context switch is a mechanism by which the OS pauses the execution of one thread and schedules another thread to run on the system. The OS can perform a context switch for various reasons, such as when a thread has finished executing, when a thread has been running for too long, or when a higher-priority thread enters the scheduling queue. In the event of a page fault, the operating system must fetch data from the disk, which is a very slow operation that leaves the CPU idle. To better use the CPU, the operating system will typically select another thread to run and perform a context switch.

Based on this observation, we implement the covert channel using the following approach: the trojan and the spy process share two pages. The trojan stores one page on the disk and the other in memory, depending on the bit being encoded. The spy process is multi-threaded, with each thread accessing one of the pages. The transmitted bit is determined by observing the order in which the threads execute.

The rest of the paper is organized as follows. The next section presents some background information on context switch, virtual memory, and covert channels. Section 3 presents the threat model of our attack. Section 4 describes the construction of our attack. We present our results in Section 5, and finally, we present some concluding remarks in section 6.

\section{Background}
\vspace{3mm} 

\subsection{Covert Channels}

Covert channels are a way for two entities to exchange information secretly without being detected by third parties. They consist of two malicious processes: a trojan (sender) and a spy (receiver). The trojan encodes information by altering shared resources, and the spy decodes the transmitted data by detecting these changes. Covert channels are usually classified into two categories: storage channels and timing channels.

\textbf{Storage Channels}: Covert storage channels transmit information by reading from and writing to a storage location. For example, a file system storage channel can be used, where the trojan modulates certain file attributes (such as the size, permissions, and access timestamps) to encode data. The spy process can then infer the encoded bit by monitoring changes to these attributes.\cite{p}

\textbf{Timing Channels}: Timing-based covert channels exploit changes in shared resources that can be observed by measuring the execution time of specific software operations. These channels often rely on changes in microarchitectural state that can be detected through differences in execution times. Cache covert channels are a common example of timing channels. For instance, when transmitting a \textit{1}, the trojan may flush data from the cache. After a pre-determined amount of time, the spy can observe the change in access latency to the memory location and infer the encoded bit. \cite{m}

\subsection{Context switch}
The scheduler is a crucial component of the operating system responsible for managing and allocating system resources, such as CPU time and memory, among the various processes and threads running on the system. 

To prevent a process from monopolizing the system resources, the OS scheduler switches the CPU from one process to another to effectively utilize the available resources. This mechanism is known as a context switch. The scheduler uses various algorithms and techniques to determine the next process or thread to run. Some of the factors that may be taken into account by the scheduler include the priority of the process or thread, the amount of time it has spent waiting to run, and the resources it requires. During a context switch, the OS temporarily suspends the execution of one process, saves its state( including registers, program counter, page table mappings, etc.), and restores the saved state of the next process or thread that will be scheduled to run. 

When a process requests an I/O operation, such as reading or writing to a file on a disk with a slow access time, the operating system typically pre-empts the requesting process and performs a context switch.

\subsection{Virtual Memory and Page Cache}
Virtual memory is a memory management mechanism that allows
secondary memory to be addressed as though it were a part of the
main memory. The motivation behind virtual memory is twofold.
It offers protection and allows for easier memory management.
The basic idea behind virtual memory is that each program runs
in its own private virtual address space, which is broken up into
chunks called pages. These pages are mapped onto physical memory. Virtual memory implements this translation from a program’s
address space to physical memory through page tables that reside in memory. Page tables store the mapping of virtual addresses onto physical addresses and are maintained on a per-process basis.

Virtual memory also allows an application's memory footprint to exceed the system’s physical memory size. It achieves this by managing two levels of the memory hierarchy i.e., physical memory and secondary storage. The operating system treats a portion of the disk as an extension of the system's physical memory. When the physical memory gets exhausted, the operating system moves some of the data to the disk, as determined by the page eviction algorithm it follows. The page tables are updated accordingly to reflect these changes. 

When a program references a virtual address that is not mapped to the physical memory, the CPU traps to the operating system. This event is known as a page fault, during which the operating system takes over and fetches the data from the disk. But as we will soon see, not all page faults require disk access.

Page cache is a software cache that has been incorporated into all major operating systems to store the working set of the processes running on the system. The operating system uses the page cache to store frequently used pages from secondary storage, such as a hard disk or solid-state drive, in main memory to evade otherwise expensive disk operations.

A \textit{soft page fault} happens when a process asks for data that is already in the page cache. This is different from a \textit{hard page fault}, where the process requests data that is not in main memory and must be retrieved from the disk, which is slower and requires a context switch. A soft page fault only involves updating the page table mapping with the page from the page cache and does not require a context switch.

\section{Threat Model}

This threat model assumes an adversary capable of running a trojan and a spy process on the same machine. An application has a trojan with access to sensitive data such as passwords and private keys. This attack requires the system to
\begin{itemize}
    \item provide a way for processes to have read access to a shared memory region (such as shared libraries or \textit{mmap}\cite{q} with the \textit{MAP\_PRIVATE} option in Linux).
    \item provide a utility that sets the CPU affinity of a process (\textit{taskset} in Linux).
    \item optimally arrange data in the memory hierarchy based on access patterns indicated by the process (using the \textit{posix\_fadvise} function in Linux).
    \item perform a context switch on a hard page fault.
\end{itemize}

\vspace{5mm} 

\section{Covert Channel Construction}

\subsection{Algorithm}

This attack fundamentally relies on the residency of a page. The trojan and spy are depicted in pseudo-code in Figure 2, with the trojan being a single-threaded process and the spy being multi-threaded. For the attack to be successful, the spy process must be restricted to a single core, and the operating system  must alternate between the spy threads when a page fault occurs. To achieve this, the \textit{taskset} utility is used to pin the spy process to a single core. It involves two pages, \textit{P1} and \textit{P2}, shared between the two processes, pointed to by \textit{addr\_1} and \textit{addr\_2}. Each iteration of the attack involves communicating a single bit and is comprised of three phases:
\vspace{3mm} 

\textit{Pre-attack setup}. The processes are synchronized by sampling \textit{clock\_gettime()}. To emulate a shared library, we share memory between the trojan and the spy process with an mmap of the shared files as \textit{MAP\_PRIVATE}, which maps the files as \textit{copy-on-write}.

\noindent
\framebox[0.97\linewidth]{%
\begin{minipage}{0.95\linewidth}
\begin{multicols}{2}
\begin{lstlisting}[language=c]
// Sender 

for every_bit {
	sync_with_receiver()
	/* mmap (MAP_PRIVATE) */
	load_to_disk(addr_1)
	load_to_disk(addr_2)
	/* posix_fadvise(POSIX_FADV_DONTNEED) */
	clear_disk_cache()
	if(bit)
		load_to_mem(addr_2)
	else
		load_to_mem(addr_1)
}















\end{lstlisting}
\columnbreak
\begin{lstlisting}[language=c]
// Receiver

for every_bit {
sync_with_sender()
t1.run()
t2.run()
join(t1)
join(t2)
print(bit)
}

thread t1() {
    fence()
    1oad = *(addr_1)
    fence()
    bit = 1
}

thread t2() {
    fence()
    load = *(addr_2)
    fence()
    bit = 0
}

\end{lstlisting}
\end{multicols}
\end{minipage}
}

\vspace{5mm} 

\textbf{Figure 2: Algorithm for Sender and Receiver}
\vspace*{3mm}

 We also noticed that the files are mapped onto the page cache by the kernel. Future access to these mapped regions will result in a soft page fault. To force hard page faults in the spy process, the trojan must evict these pages from the page cache. We experimented with the \textit{posix\_fadvise} system call to tackle this problem. It allows us to hint to the kernel about the pattern in which we access a file in the future. More specifically, the advice \textit{POSIX\_FADV\_DONTNEED} indicates that the process will not be accessing the file shortly, allowing the kernel to evict it from the page cache to improve performance. 

\vspace{3mm} 

\textit{Encoding data}. The Trojan can encode data by manipulating the residency of pages, which are initially stored on the disk. 
To encode a \textit{1}, the trojan accesses the page pointed to by \textit{addr\_2}, causing it to be brought into memory. Subsequent accesses to the same page will not result in a page fault. Similarly, to encode a \textit{0}, the trojan accesses the page pointed to by \textit{addr\_1}. Accessing this page will also bring it into memory.

\vspace{3mm} 

\textit{Decoding data}. In the final phase of the attack, after a predetermined amount of time has passed since the encoding phase (known as the synchronization period), the threads $t1$ and $t2$ in the spy process access the data from \textit{addr\_1} and \textit{addr\_2}, respectively. If $t1$  executes last, it means that $t1$ experienced a page fault, indicating that the page pointed to by \textit{addr\_2} is located on the disk and the trojan encoded a \textit{0}. If $t2$ executes last, it means that $t2$ experienced a page fault, indicating that the page pointed to by \textit{addr\_2} is located on the disk and the trojan encoded a \textit{1}. As shown in figure 3, the trojan can control the order in which the threads in the spy process execute.

\vspace{5mm} 

\shiftleft{10pt}{
\includegraphics[height=0.175\textheight]{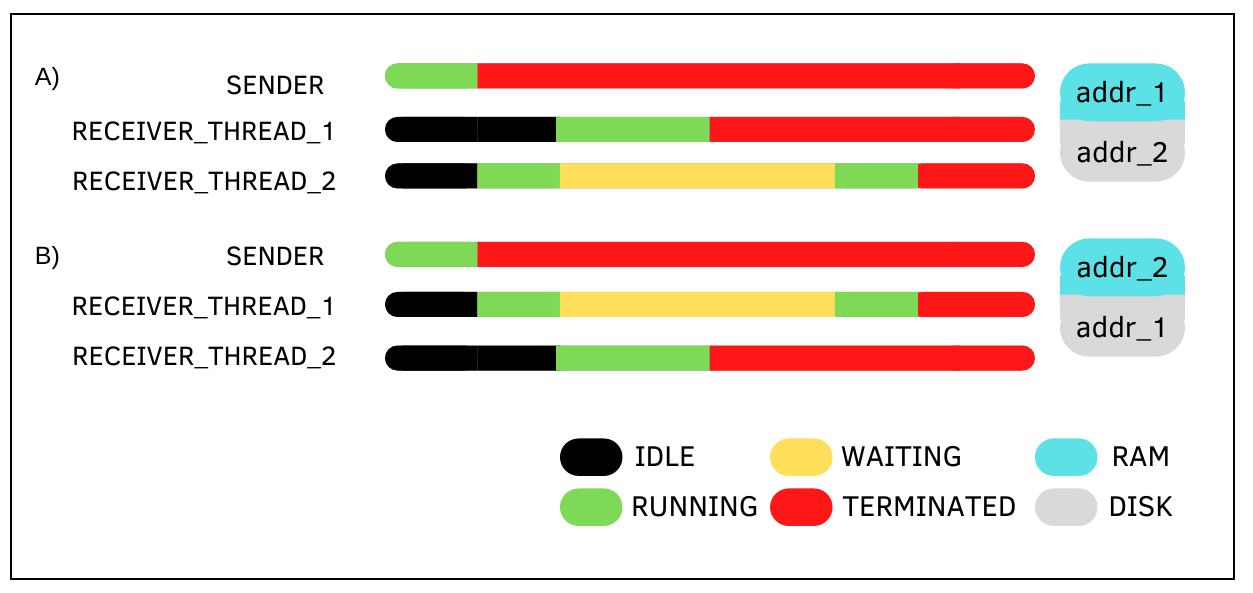}
}

\textbf{Figure 3: Timing of the colluding processes. \\ (A) Trojan transmits a \textit{0}. (B) Trojan transmits a \textit{1}. }

\begin{figure}
     \centering
     \begin{subfigure}[b]{0.4\textwidth}
         \centering
         \includegraphics[width=0.8\textwidth]{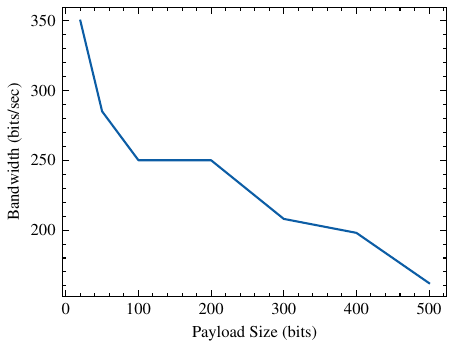}
         \caption{Bandwidth vs Payload Size}
         \label{fig:Bandwidth vs Payload Size}
     \end{subfigure}
     \hfill
     \begin{subfigure}[b]{0.4\textwidth}
         \centering
         \includegraphics[width=0.8\textwidth]{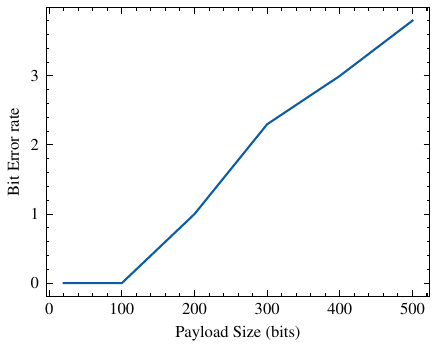}
         \caption{Bit Error Rate vs Payload Size}
         \label{fig: Bit Error Rate vs Payload Size}
     \end{subfigure}
\vspace{3mm}

\textbf{Figure 4: (A) Bandwidth vs Payload Size (B) Bit Error Rate vs Payload Size}
\end{figure}



\subsection{Sensitivity to Access Pattern} Every iteration of the spy process involves loading two pages, and the order in which these pages are accessed can affect the state of the page cache, affecting the channel's accuracy. To determine an optimal access pattern, we systematically generated sequences of two pages, with each sequence being separated by a certain number of pages, the page gap (\textit{M}), i.e, P1, P2 for \textit{bit 1}, \textit{P1 + M}, \textit{P2 + M} for \textit{bit 2}, \textit{P1 + 2M}, \textit{P2 + 2M} for \textit{bit 3}, etc. We repeat this experiment 7 times for \textit{M = \{4, 8, 16, 32, 64, 128, 256\}} for different payload sizes and measure the bit error rate in each case. A lower error rate for an access pattern indicates that it is more efficient in tricking the kernel into not loading data into the page cache. Figure 5(a) shows how the error rate varies with the page gap for a payload size of 50 bits. For this channel, we choose the value of M that results in the lowest error rate for different payload sizes. After accessing the last page entry in the shared memory, the trojan and spy wrap around to the beginning of the shared memory. We considered a scenario in which the trojan and spy have access to a shared memory of 32 MB with read-only permissions. The impact of different sizes of shared memory will be examined in Section 5.1. Table 2 shows the chosen values of M for different payload sizes.
\vspace{3mm}

\scalebox{0.92}{
\begin{tabular}{|c|c|}
\hline
\textbf{Payload Size (bits) } & \textbf{Page Gap (M)} \\
\hline
20 & 128 \\
\hline
50 & 128 \\
\hline
100 & 64 \\
\hline
200 & 64 \\
\hline
300 & 32 \\
\hline
400 & 16 \\
\hline
500 & 16 \\
\hline
\end{tabular}
}

\vspace{3mm}

\textbf{Table 2: Optimal page gap for different payload sizes}

\vspace{3mm}

\begin{figure}
     \centering
     \begin{subfigure}[b]{0.4\textwidth}
         \centering
         \includegraphics[width=0.8\textwidth]{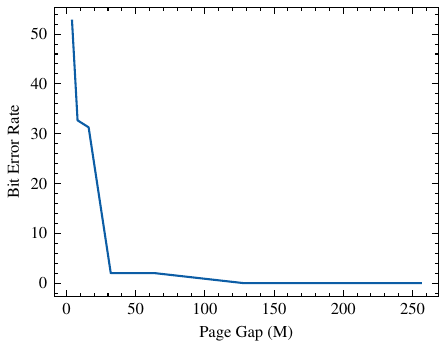}
         \caption{Bit Error Rate vs Page Gap (M)}
         \label{fig: Bit Error Rate vs Page Gap (M)}
     \end{subfigure}
     \hfill
     \begin{subfigure}[b]{0.4\textwidth}
         \centering
         \includegraphics[width=0.8\textwidth]{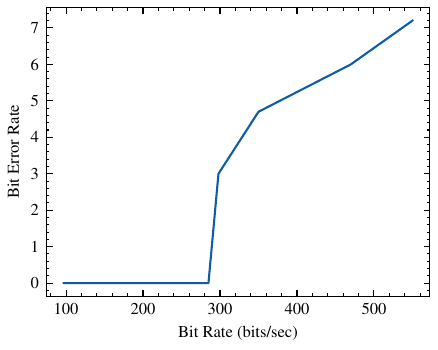}
         \caption{Bit Error Rate vs Bit Rate}
         \label{fig: Bit Error Rate vs Bit Rate}
     \end{subfigure}
\vspace{3mm}

\textbf{\textbf{Figure 5: (A) Bit Error Rate vs Page Gap (M) (B) it Error Rate vs Bit Rate}}
\end{figure}

\vspace{-3mm}

\section{Results}
 We run our experiments on an eight-core AMD Ryzen 7 5800H CPU running Ubuntu 20.04.4 LTS. The attack was also successfully launched on an Intel Core i7-8750 CPU running Ubuntu 20.04 LTS. 

We pin the spy process to a single core to ensure that the threads in the spy process compete for the same core. 
\subsection{Sensitivity to Payload and Shared Memory Size}
 Figures 4(a) and 4(b) show how the payload size affects the bandwidth and accuracy of the channel, respectively. As the payload size increases, the accuracy of the channel decreases. This is likely due to the fact that when transmitting large payloads, the sender has to wrap around to the beginning of the shared memory multiple times, which leads to the accessed pages being stored in the page cache and reducing the performance of our channel. Increasing the size of shared memory (this can be easily done by the colluding processes by mapping a shared memory region to a large shared library or a sequence of smaller libraries through the \textit{mmap} function) would prevent the sender from wrapping around as often, which would reduce the likelihood of caching the data and improve the accuracy of the channel. Table 3 demonstrates how the performance changes based on the size of the shared memory for a payload size of 100 bits.

\vspace{3mm}

\begin{tabular}{|c|c|}
\hline
\textbf{Shared Memory Size} & \textbf{Bit Error Rate} \\
\hline
32MB & 1\% \\
\hline
16MB & 4\% \\
\hline
8MB & 9\% \\
\hline
4MB & 13\% \\
\hline
2MB & 21\% \\
\hline
1MB & 32\% \\
\hline
\end{tabular}

\vspace{3mm}

\textbf{Table 3: Bit Error Rate vs Shared Memory Size}

\vspace{3mm}

\subsection{Sensitivity to transmission rate} 
Figure 5(b) shows the bit error rate vs the transmission rate for this channel. To determine the bit error rate at different transmission rates, the synchronization period is modified while keeping the payload size constant. As the bit rate increases, the error rate also increases. This is because the channel requires rather expensive operations (memory loads for the sender and disk loads for the receiver) to be completed within the synchronization window. If this window is too short, it can result in incorrect inferences and a higher error rate.

\section{Conclusions}
This paper introduces a new mechanism for covert communication that does not rely on timers. This method can be used on many different types of systems and does not rely on specific system features. However, it relies on page faults and disk operations, which are slow and limit the data transfer rate of this method. It is, therefore, slower than other state-of-the-art techniques.

\end{document}